\documentclass{osa-article}

\journal{osajournal}

\articletype{Research Article}
\usepackage{xcolor}
\usepackage{lineno}
\usepackage{graphicx}
\usepackage{amsmath}
\usepackage{siunitx}
\usepackage{caption}

\newcommand{\beginsupplement}{%
        \setcounter{section}{0}
        \renewcommand{\thesection}{S\arabic{section}} 
        \setcounter{figure}{0}
        \renewcommand{\thefigure}{S\arabic{figure}}%
     }
\usepackage{float}
\usepackage{url}

\begin{document}

\title{Using electrical resistance asymmetries to infer the geometric shapes of foundry patterned nanophotonic structures}

\author{Vinita Mittal and Krishna C. Balram}

\address{Quantum Engineering Technology Labs and Department of Electrical and Electronic Engineering, University of Bristol, Woodland Road, Bristol, United Kingdom BS8 1UB}

\email{vinita.mittal@bristol.ac.uk; krishna.coimbatorebalram@bristol.ac.uk} 



\begin{abstract*}
While silicon photonics has leveraged the nanofabrication tools and techniques from the microelectronics industry, it has also inherited the metrological methods from the same. Photonics fabrication is inherently different from microelectronics in its intrinsic sensitivity to 3D shape and geometry, especially in a high-index contrast platform like silicon-on-insulator. In this work, we show that electrical resistance measurements can in principle be used to infer the geometry of such nanophotonic structures and reconstruct the micro-loading curves of foundry etch processes. We implement our ideas to infer 3D geometries from a standard silicon photonics foundry and discuss some of the potential sources of error that need to be calibrated out to improve the reconstruction accuracy.
\end{abstract*}

\section{Introduction}

In the past decade, foundry silicon photonics has successfully leveraged complementary metal oxide semiconductor (CMOS) technology to revolutionize modern optics with a wide-ranging impact on diverse areas, spanning telecommunications \cite{reed2010silicon, witzens2018high} and sensing \cite{steglich2019optical} to quantum computation \cite{bonneau2016silicon, harris2016large}. The scale and complexity of devices that researchers can engineer using modern silicon photonics foundries was unthinkable even a few years ago. Today, more than 30 fabrication steps are performed on a 300 mm diameter silicon wafer to manufacture millions of high-performance devices at a wafer scale with high yield. While silicon photonics has leveraged the nanofabrication tools and techniques from the microelectronics industry, it has also inherited the metrological methods from the same. This is a key limitation in photonics fabrication, because it is inherently different from microelectronics in its intrinsic sensitivity to 3D shape and geometry. For photonics processes, the 3D geometry of the fabricated structure is critical to wave confinement, which requires correcting for the foundry's micro-loading effect or aspect-ratio dependent etching, where the etch depth is dependent on the pattern fill fraction \cite{rangelow2003critical}. This problem is further exacerbated in the modern era of computationally designed photonic structures, which rely on advances in numerical methods and computational hardware to design structures that are optimal at mapping a given input field distribution to a specified output field pattern \cite{wetzstein2020inference, molesky2018inverse}. All such inverse design methods assume that the generated complex 3D shapes (refractive index profiles) can be fabricated with high reliability and reproducibility. This is not true in practice, especially when the fabricated structures have multi-level etch depths (partial etches). Pushing metrology techniques is therefore critical, and developing novel metrological methods that allow one to infer the shapes of foundry-patterned nanophotonic devices, without resorting to destructive cross-sectional scanning electron microscopy (SEMs) or time-consuming atomic force microscopy (AFM) scans, are key not only for process repeatability and control, but also enabling the next generation of inverse-designed optimal nanophotonic devices. 

Accuracy in feature dimensions and their repeatability across a wafer are the two key limiting factors contributing to the overall design yield of any fabrication process. In a photonics foundry, a number of copies of the same device are usually made to estimate the fabrication variations so that appropriate measures such as thermal tuning can be implemented to compensate for manufacturing errors \cite{zortman2010silicon, giewont2019300}. It not only restricts the available mask area, hence increasing both the cost of materials and processing time, but also puts additional constraints on power consumption if external compensation strategies are adopted. Therefore, minimising these errors by accurate estimation of 3D geometries of the devices is crucial. Some of the leading silicon photonics foundry service providers report a dimensional tolerance of $\approx$ 14\% ($\pm$ 10 nm for a 70 nm etch) \cite{Cornerstone}, which makes it challenging to design resonant structures accurately. To compensate for these manufacturing errors, various techniques such as topological feature optimisation \cite{jensen2011topology}, resolution enhancement \cite{orlando2016opc} and proximity error correction \cite{celo2017optical} are applied at the photonic mask design level. However, these methods only correct for the 2D shapes, similar to what is standard in CMOS electronics foundries and not for the overall 3D shape, which is necessary for photonics. Predictive statistical models such as Monte Carlo simulation methods can quantify the impact of layout-dependent correlated manufacturing variations on the performance of photonics devices, for instance, by measuring the spectral response from a large number of identical devices and quantifying the manufacturing errors \cite{lu2017performance, rausch2019monte}. Similarly, optical methods based on measuring transmission spectra from different cross-sections of Mach- Zehnder interferometer devices have been used to infer the geometrical variations in silicon waveguides \cite{oton2016silicon}. Not only are these iterative methods tedious and time consuming, they inevitably require some geometrical compensation to account for the variations due to local pattern density changes \cite{lu2017performance}.

In current foundry processes, at the mask level a metrology box (of area $\approx$ 1 mm$^2$) is usually drawn on each mask layer of the device to evaluate errors in the etch depth using ellipsometry after each subsequent process step. Since the sensitivity of ellipsometry is limited by the available spot size, it can not provide an accurate estimate for small (sub-micron) and closely spaced features such as grating couplers and  ring resonators, where the gap between a bus waveguide and the ring is $\approx$ 200 nm. Therefore, SEMs are commonly used for assessing the surface profile of these devices. Both the (partial) etch depth and sidewall angle profiles of the patterned structures are estimated from a focused ion beam (FIB) cross-section of the sample using a critical dimension SEM (CD-SEM). In practice, one needs to be careful in relying on SEM metrology of device cross-sections as it is sensitive to various parameters like edge effects, sample charging, beam parameters, tool calibration and the electron beam-specimen interaction \cite{postek2013does}. Moreover, it requires multiple measurements as the etch profile of closely spaced small features is not similar to that of large features, due to the micro- loading effect observed in reactive ion etching. To the best of our knowledge, no foundry process currently includes this important information in their process design kit (PDK). Hence, there is a real need for newer, non-destructive techniques to calibrate on-chip 3D photonic geometries. In a high index contrast platform like silicon-on-insulator (SOI), small changes in waveguide geometry can lead to significant changes in the mode index, as we discuss in detail below.

\begin{figure}[H]
    \centering\includegraphics[width=1\textwidth, trim = 0.5cm 5cm 0cm 5cm, clip]{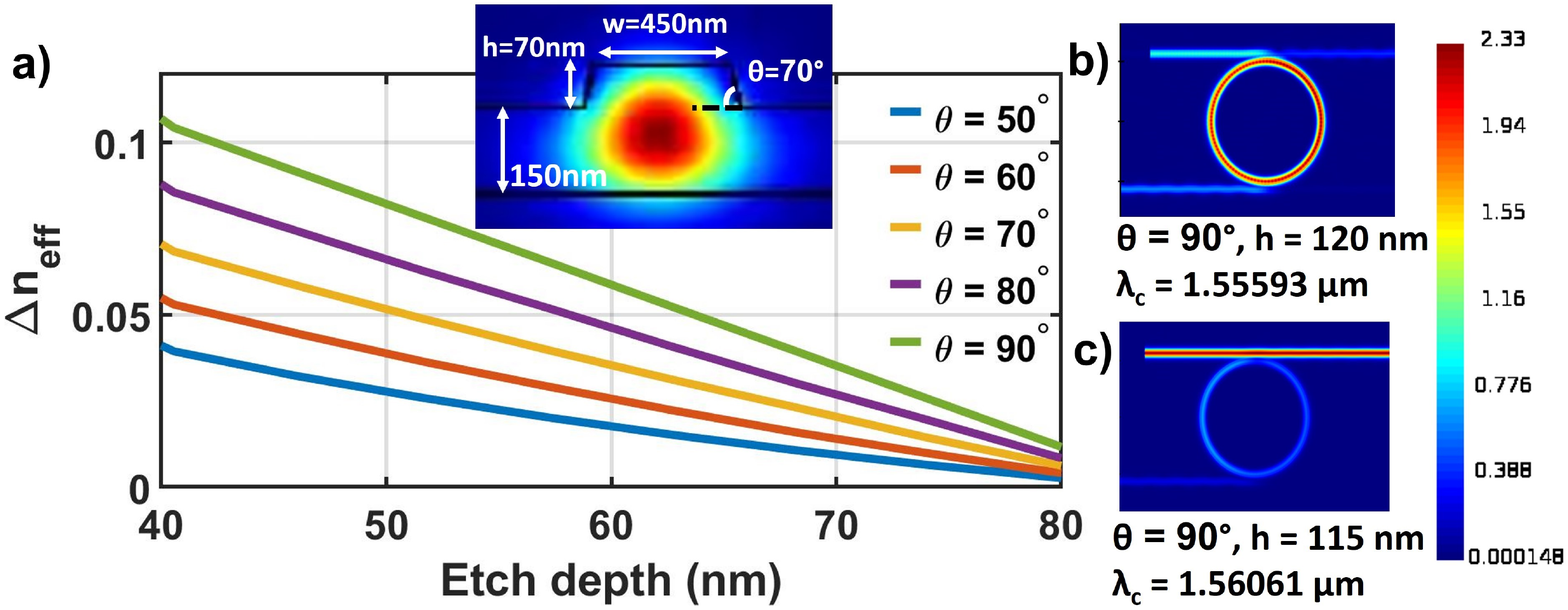}
    \caption{(a) Numerical FDE (Lumerical) simulation of the mode index ($n_{eff}$) of the fundamental TE mode of a 220 nm SOI rib waveguide showing the change in effective refractive index (${\Delta}n_{eff}$) with respect to the slope angle ($\theta$) and etch depth ($h$) at $\lambda$ = 1550 nm. The inset shows the normalized electric field distribution for a sidewall angle of 70\textdegree. (b) and (c) FDTD simulations show the sensitivity of photonic device performance to 3D geometry by calculating the resonance wavelength of a rib micro-ring resonator (silicon thickness = 220 nm, waveguide width = 450 nm, $\theta$ = 90\textdegree and a coupling gap of 200 nm), with etch depth (b) $h$ = 120 nm (slab thickness = 100 nm) and (c) $h$ = 115 nm (slab thickness = 105 nm), respectively. The small change in etch depth changes the resonance wavelength by $\approx$ 4.7 nm. The field plots in (b) and (c) are both evaluated at 1.55593 \si{\um}}
    \label{fig 1:Lumerical}
\end{figure}

To get a sense of the problem, we consider the simple case of a standard SOI rib waveguide (silicon thickness = 220 nm) and the dependence of the mode index ($n_{eff}$) on waveguide geometry. Fig.1(a, inset) shows the normalized electric field distribution of the fundamental transverse electric (TE) mode of an SOI waveguide with slab thickness of 150 nm, rib height, $h$ = 70 nm and sidewall angle ($\theta$) 70\textdegree, at a wavelength of $1550$ nm. The change in the mode index due to geometry ($\Delta$ $n_{eff}$), as a function of $h$ and $\theta$ is plotted in Fig.1(a), calculated using a numerical solver (Lumerical Finite Difference Eigenmode (FDE)). As can be seen, small changes in $h$ and $\theta$ lead to relatively  large changes in $n_{eff}$, which is understandable given the tight mode confinement in the SOI platform. This is best illustrated by Fig.1(b) and (c), where an example \cite{lumerical} from Lumerical MODE 2.5D variational Finite Difference Time Domain (FDTD) solver on a partially etched SOI ring resonator is simulated for rib height (etch depth) of 120 nm and 115 nm, and their resonance centre wavelength is compared. The ring radius is 5 \si{\um}, waveguide width is 450 nm, slab thickness is 100 nm and gap between the ring and waveguide is 200 nm. As can be seen from Fig.1(b) and (c), a 5 nm change in etch depth led to a shift in central wavelength by 4.7 nm. As the variation in the etch depth increases, along with the cumulative effect of variation in sidewall angle, it leads to significant change in the device characteristics.

In this work, we use the geometrical dependence of electrical resistance measurements to infer the shapes of foundry-patterned nanophotonic devices. In particular, we show that differential resistance measurements can eliminate many of the systematic sources of error and uniquely serve as a probe of the geometrical shape of the patterned device. To the best of our knowledge, we believe this is the first attempt to use precision electrical measurements as a non-destructive probe for 3D nanofabrication metrology and process control, especially for integrated photonics. With that in mind, we also list some other sources of error that need to be calibrated out to improve the overall accuracy of our method.

\section{Working principle: geometry and resistance asymmetry}

\begin{figure}[H]
    \centering\includegraphics[width=1\textwidth, trim = 0cm 5cm 0cm 5cm, clip]{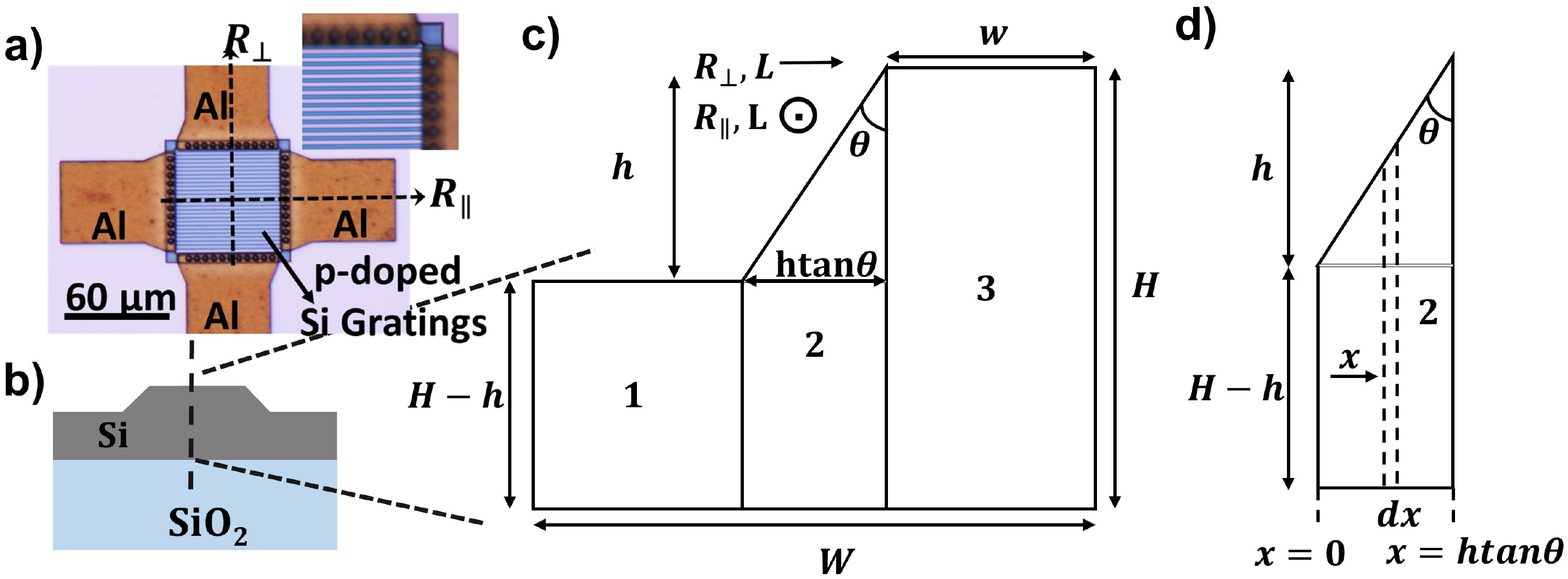}
    \caption{(a) An optical microscope image of a representative grating device indicating the parallel ($R_{||}$) and perpendicular ($R_{\perp}$) direction of measurement with respect to the grating axis, shown in a magnified image in the inset. (b) A schematic of the cross-section of a single grating element showing angled walls, which is magnified in (c) and (d) showing a unit cell of a grating sub-divided into 3 regions. The direction of length of the grating is also indicated in (c) for both parallel and perpendicular direction of measurement. (d) shows a zoom-in on region 2, showing the elemental resistance unit used for integral evaluation, see text for details.}
     \label{fig 2: Unit cell theory}
\end{figure}

Consider a standard single mode p-doped SOI rib waveguide with the total thickness, $H$ = 220 nm and etch depth, $h$ = 70 nm. Grating-like structures are designed with various periods ($\Lambda$) and lengths ($L_g$) and implemented in an active silicon photonics foundry platform, detailed further in Section 3. The main goal of our work is to try and infer the partial etch depth ($h$) and sidewall angle ($\theta$) of a fabricated device through a non-destructive electrical measurement. The devices are designed to have a nominal etch depth of 70 nm independent of $\Lambda$, but due to the micro-loading effect discussed above, the etch depth depends (quite strongly) on $\Lambda$ and we would like to reconstruct this dependence from electrical measurements of the device resistance.

Our main insight is that device resistance is usually determined by geometry, and in the right operating regime, this relationship can be inverted to reconstruct the geometry of a fabricated device. An SOI photonics process provides a good test platform for these ideas because the current flow is limited to the silicon device layer and it is the silicon layer geometry that we are primarily interested in reconstructing. Both the partial etch depth ($h$) and the side wall angle ($\theta$) can be estimated by measuring the device resistance in both parallel (\(\text{R}_{||}\)) and perpendicular (\(\text{R}_{\perp}\)) directions, with respect to the patterned gratings, as shown in the microscope image of a representative fabricated device in Fig.2(a). We label the parallel direction to be along the length (long axis) of the gratings and the perpendicular direction is defined along the width (short axis) of the gratings. Fig.2(b) shows a schematic cross-section of a single grating element, from which a unit cell is drawn, and described in Fig.2(c). The unit cell has three regions numbered 1, 2 and 3. Region 1 describes the slab height which equals $H$-$h$. The total width ($W$) of the unit cell is one half of the grating period ($\Lambda$) and the width of the region 3 is denoted by $w$. The total resistance measured along parallel (\(\text{R}_{||}\)) and perpendicular (\(\text{R}_{\perp}\)) direction of a grating can be analytically calculated as follows (derivations are provided in the supplementary information(SI), section S1):

\begin{equation}
   \mathrm R_{||}\ = \frac{\rho L_g}{(HW - hW + hw + \frac{1}{2}h^2 \mathrm{tan}\theta)}.
\end{equation}
\begin{equation}
    \mathrm R_\perp\ = \frac{\rho}{L_g} \left (\frac{W-h \mathrm{tan}\theta-w}{H-h} + tan\theta  \ln\left(\frac{H}{H-h}\right) +\frac{w}{H}\right). 
\end{equation}
where
     \begin{equation}
     W= \frac{\Lambda}{2}
\end{equation}
 \begin{equation}
      w= {\Lambda}*\frac{FF}{2}
\end{equation}

where FF is the grating fill fraction, as measured from a top view optical microscope or SEM image.

Each device has N grating elements, denoted by $N_{\Lambda}$ which is given by:
\begin{equation}
      N_{\Lambda} = \frac{2L_g}{\Lambda} 
\end{equation}
We define the number of periods in the mask \(\text{N}_{pd}\) as follows:
\begin{equation}
     \mathrm N_{pd} = \frac{N_{\Lambda}}{2} 
\end{equation}
with the factor of 2 accounting for the unit cell being half the period.

When the resistance measurement is performed along the parallel direction (\(\text{R}_{||}\)) as shown in Fig.2(c), each of the $N_{\Lambda}$ grating elements are connected in parallel whereas for the measurement along the perpendicular direction (\(\text{R}_{\perp}\)), each $N_{\Lambda}$ grating elements are connected in series. Taking this into account, the overall contributions from \(\text{R}_{||}\) and \(\text{R}_{\perp}\) from equations (1) and (2) can be written as:
\begin{equation}
    \mathrm R_{||}\ = \frac{\rho L_g}{N_{\Lambda} (HW - hW + hw + \frac{1}{2}h^2 \mathrm{tan}\theta)}
\end{equation}

\begin{equation}
   \mathrm R_\perp\ = \frac{N_{\Lambda} {\rho}}{L_g} \left (\frac{W-h \mathrm{tan}\theta-w}{H-h} + \mathrm{tan}\theta  \ln\left(\frac{H}{H-h}\right) +\frac{w}{H}\right) \\
\end{equation}
The resistance asymmetry (${\Delta}$R) is then defined as:
\begin{equation}
    \Delta \mathrm R= \mathrm R_\perp\ - \mathrm R_{||}\
\end{equation}

Ideally, two equations are obtained with three unknowns $h$, $w$ and $\theta$ given that $H$ = 220 nm. The width of the fabricated structures ($w$), or more precisely the fill fraction (FF), can be obtained non-destructively from an optical microscope or SEM inspection. Therefore, in practice it requires solving two unknowns $h$ and $\theta$, which can be done uniquely for each device.

Using ${\Delta}$R to estimate the geometry relies on some key underlying assumptions that we will revisit later in this work. We assume that the (top and bottom) oxide resistivity is infinite and all the current flows through the silicon device layer. We further assume that the sample resistivity ($\rho$) is independent of geometry. More precisely ($\rho_\perp = \rho_{||}$). We also assume that $\rho$ is independent of the grating length. In particular, there are no shadowing and edge effects during the ion implantation process. We assume that the contact resistance is identical in the two directions, and therefore cancels out for ${\Delta}$R. Finally, we assume that the silicon layer is doped sufficiently high so that the metal-semiconductor contact is ohmic in nature and the current flows uniformly through the device cross-section. While some of these assumptions might appear a bit idealized, we show below that there is an operating regime in which they are broadly satisfied, and device resistance can be used to infer geometry.

\section{Resistance asymmetry characterization: electrical and FIB x-section}

A representative device typically consists of partially etched gratings on a 220 nm silicon device layer, with a rib height of 70 nm, defined by the grating etch step in an active 220 nm silicon photonics foundry process \cite{littlejohns2020cornerstone}. The grating region is P-doped (Boron, doping concentration of $\approx$ 3.8x10$^{17}$ cm$^{-3}$ implanted at an angle of 7\textdegree). The contact region is $P^{++}$doped (Boron doping concentration of $\approx$ 1x10$^{20}$ cm$^{-3}$) to ensure ohmic contacts. The devices were fabricated using the actives process of a UK photonics foundry, Cornerstone \cite{littlejohns2020cornerstone}. The grating period ($\Lambda$) in different test structures was varied from 500 nm to 2 \si{\um} with an interval of 100 nm and the device length ($L_g$) was varied from 20 \si{\um} to 60 \si{\um} with an interval of 10 \si{\um}. Four aluminium (Al) pads of thicknesses 1.6 \si{\um} each were fabricated on 4 sides of the gratings, and the $P^{++}$ doped silicon layer was contacted using Al plugs. A microscope image of the fabricated device is shown in Fig.2(a). Two tungsten needle probes of diameter 12.5 \si{\um} were used to measure the resistance across the device in both parallel and perpendicular directions of the gratings, as indicated in Fig.2(a), using a probe station. The resistance was recorded using a multimeter (Rhode $\&$ Schwarz, HMC 8012) and the reported values are obtained after averaging over 500 readings.

The measured difference in resistance in perpendicular and parallel direction ($\Delta$R), is plotted in Fig.3 for different grating lengths along with a nominal theoretical prediction based on equations (6) and (7). For estimating the theoretical ${\Delta}$R (purple), the intended geometrical parameters (assuming no geometrical fabrication errors) are used, $H$ = 220 nm, $h$ = 70 nm, $FF$ = 0.5, $\theta$ = 0. The silicon resistivity, $\rho$ = 7.9x10$^{-4}$ $\Omega$-m is calculated from the foundry provided boron doping concentration of 3.8x10$^{17}$ $\Omega$-m, under the assumption that all the dopants are fully activated and the resistivity of the silicon device layer is uniform post-dopant activation. This is a reasonable assumption here given the thickness of silicon device layer is small (220 nm). For reference, a second curve (green) is also shown for $\rho$ = 2.7x10$^{-3}$ $\Omega$-m to bound the data for incomplete dopant activation. 

\begin{figure}[H]
    \centering\includegraphics[width=1.0\textwidth, trim = 0.8cm 2.6cm -1.2cm 2.7cm, clip]{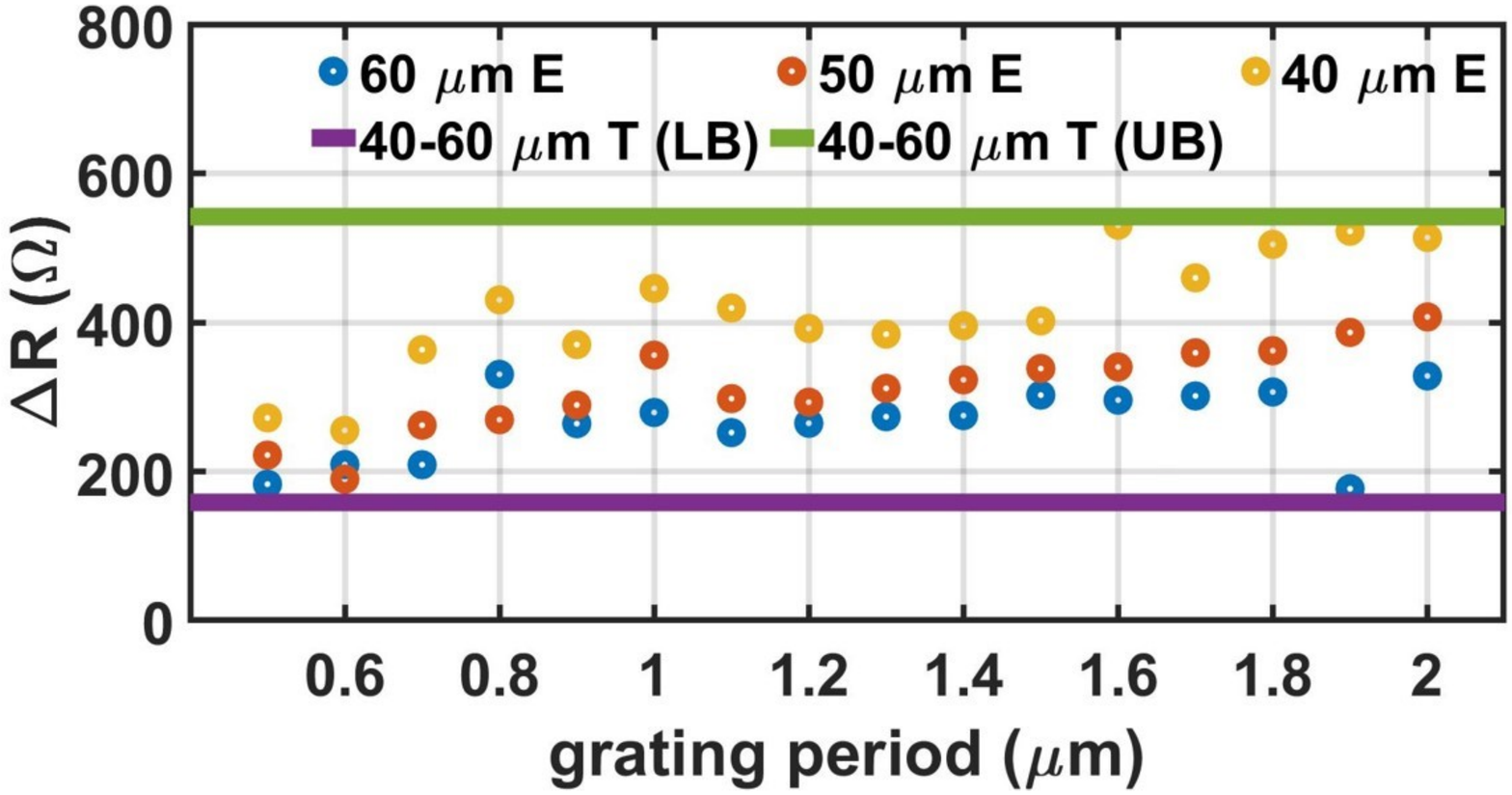}
    \caption {Resistance asymmetry ($\Delta$R) as a function of grating period for grating lengths of 60 \si{\um} (blue), 50 \si{\um} (brown)  and 40 \si{\um} (yellow) measured experimentally (E). Theoretical (T) estimates assuming perfect nanofabrication and contact symmetry for $\rho$ = 7.9x10$^{-4}$ $\Omega$-m (purple) and $\rho$ = 2.7x10$^{-3}$ $\Omega$-m (green) are also shown. The purple curve corresponds to the dopant concentration provided by the foundry and is the lower bound (LB), assuming perfect dopant activation. The green curve serves as an upper bound (UB) for the measured data. Note that the theoretical estimates for L = 40-60 \si{\um} lie on top of each other, as there should be no $L_g$ dependence for ${\Delta}$R in theory.}
    \label{fig 3: expt delta R}
\end{figure}

It is clear from the measurements that ${\Delta}$R has a significant dependence on the grating period ($\Lambda$) and grating lengths ($L_g$). From theory (eqn. 6-8), for a square grating $N_{\Lambda}$*$\Lambda$ = 2$L_g$, ${\Delta}$R should be zero. Even for non-square gratings, ${\Delta}$R should be a constant, parameterized by the silicon resistivity under the assumption that the etch parameters ($\theta$ and $h$) are independent of period. In practice, while this is very close to reality for the sidewall angle, due to the micro-loading effect discussed above, the etch depth will show a dependence on the period. The dependence of $\theta$ and $h$ on ${\Delta}$R and hence, on the microloading curve can be estimated by taking a first order differential of ${\Delta}$R with respect to $\theta$ and $h$ (shown in the SI, Fig.S1 in section S2). We find that the resistance asymmetry (${\Delta}$R) shows relatively little sensitivity to $\theta$ variations, but it is extremely sensitive to $h$ (partial etch depth) variations.

While there is some residual device to device fluctuation, from the measurements of different grating lengths, we can infer that the ${\Delta}$R monontonically increases with period and gets less steeper as we move towards longer periods. This is clearest to observe in the case of the 60 \si{\um} long devices, shown in Fig.5(a) below. We attribute the fluctuation in ${\Delta}$R to local variations in the contact resistance and dopant activation, and it serves as one of the main sources of error in our experiment, and we discuss this further below. The dependence of ${\Delta}$R on grating period hints strongly on the presence of etch micro-loading i.e., the local etch depth as a function of the period ($\Lambda$). We confirm this micro-loading effect using cross-section SEM images of these foundry fabricated devices using a dual-beam FIB (Helios NanoLab 600). Here, the sample is tilted at 52\textdegree to bring its surface perpendicular to the Ga$^+$ ion beam. Pt is deposited on the gratings to protect them from radiation damage from Ga$^+$ ions during sputtering process. First, a coarse milling of the bulk material close to the area of interest is performed by using a high beam current. Then a fine polishing cross-section is obtained by successively decreasing the beam current until the surface underneath the metal is revealed with minimum damage from the beam. A final FIB trench is then cut through the Pt surface with an optimal combination of beam energy ($\approx$ 30 keV) and beam current ($\approx$ 2.8 nA). A high resolution calibrated field effect SEM is then used to image the milled cross-section and the images are corrected for the sample tilt.

\begin{figure}[H]
\centering\includegraphics[width=1\textwidth, trim = 1.2cm 0cm 1cm 0cm, clip]{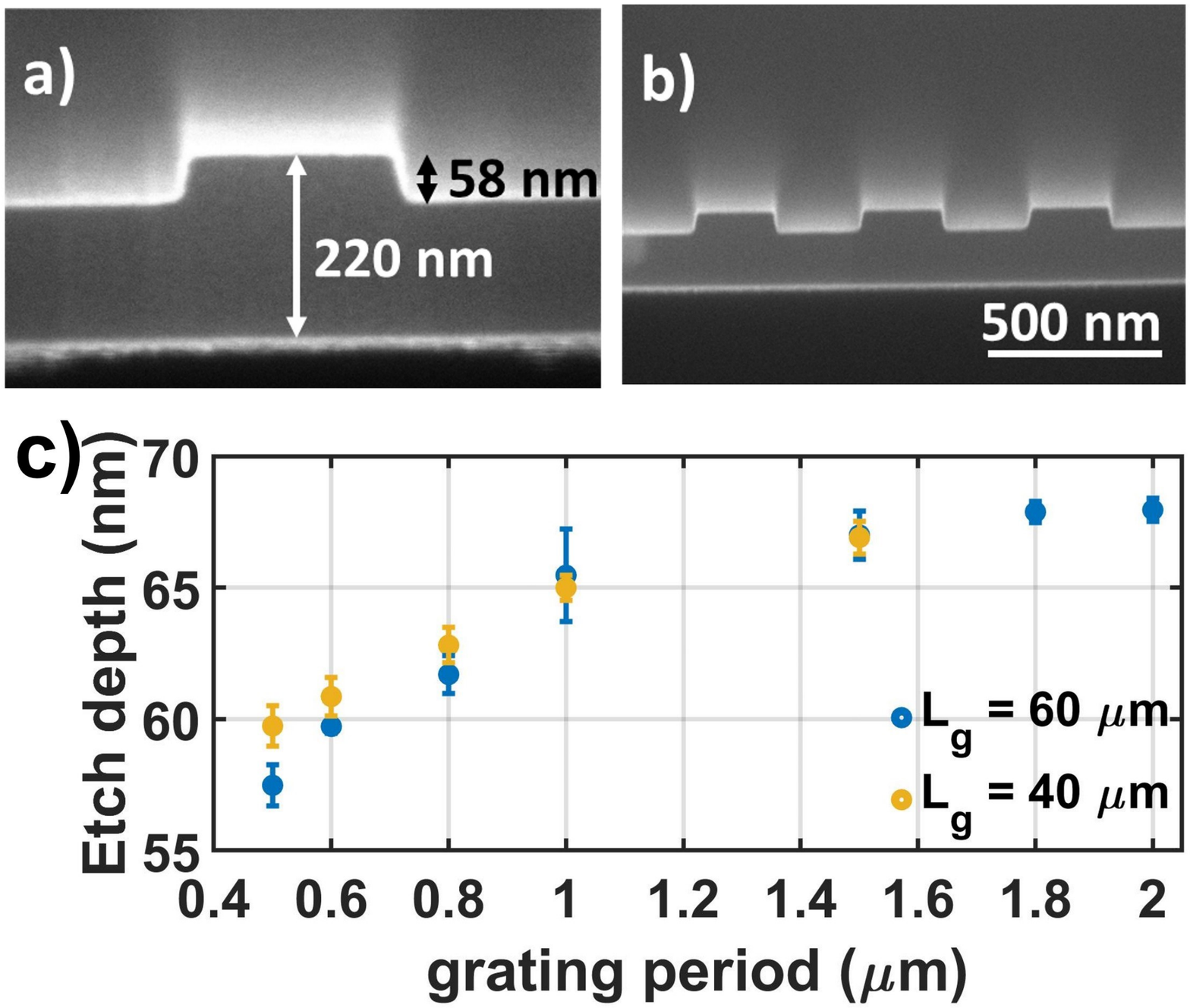}
\caption{(a) and (b) SEM image of the FIB cross-section of a representative grating device with period 0.5 \si{\um} and length 60 \si{\um}. (c) Extracted partial etch depths (from FIB x-sections) of devices with varying grating periods (nominal etch depth 70 nm) for device lengths of 60 \si{\um} (blue) and 40 \si{\um} (yellow). The error bars represent the standard deviation in measured etch depths, estimated from three different locations on the same image.}
 \label{fig 4: FIB}
\end{figure}

Fig.4(a) and 4(b) show cross-section SEM images of representative devices with period 500 nm and length 60 \si{\um}. By using a calibrated SEM, we can estimate the partial etch depth (designed to be 70 nm $\pm$ 10 nm) between the grating fingers, as in Fig.4(a) and (b). The extracted etch depth, obtained from a FIB cross-section of representative devices, as a function of grating period for devices with two different lengths ($L_g$ = 40 \si{\um} and 60 \si{\um} respectively), is shown in Fig.4(c). The measured partial etch depth shows the same trend as indicated by the ${\Delta}$R data in Fig.3 (and Fig.5(a)) more clearly. The partial etch depth monotonically increases with increasing period ($\Lambda$) and then saturates to a steady state value at large grating periods ($\approx$ 2 \si{\um}). This is roughly in line with what one would expect from standard etch micro-loading \cite{yeom2003critical}. The data shown in Fig.4(c) reconstructs the etch micro-loading curve for this particular foundry process and our main aim in this work is to determine the same from a non-destructive electrical measurement. While the qualitative agreement between the FIB measurements in Fig.4(c) and the ${\Delta}$R curve in Fig.3 (and Fig.5(a)) is clear, for the method to work as means of geometric reconstruction, we need quantitative agreement, which we explore next. In passing, we would like to note here that the foundry process shows remarkable precision and control, as evidenced by the saturating etch depth being $\approx$ 68 nm, only 2 nm off from the designed etch depth of 70 nm. 

\begin{figure}[!htbp]
\centering\includegraphics[width=1\textwidth, trim = 0.35cm 3.5cm -0.75cm 3.6cm, clip]{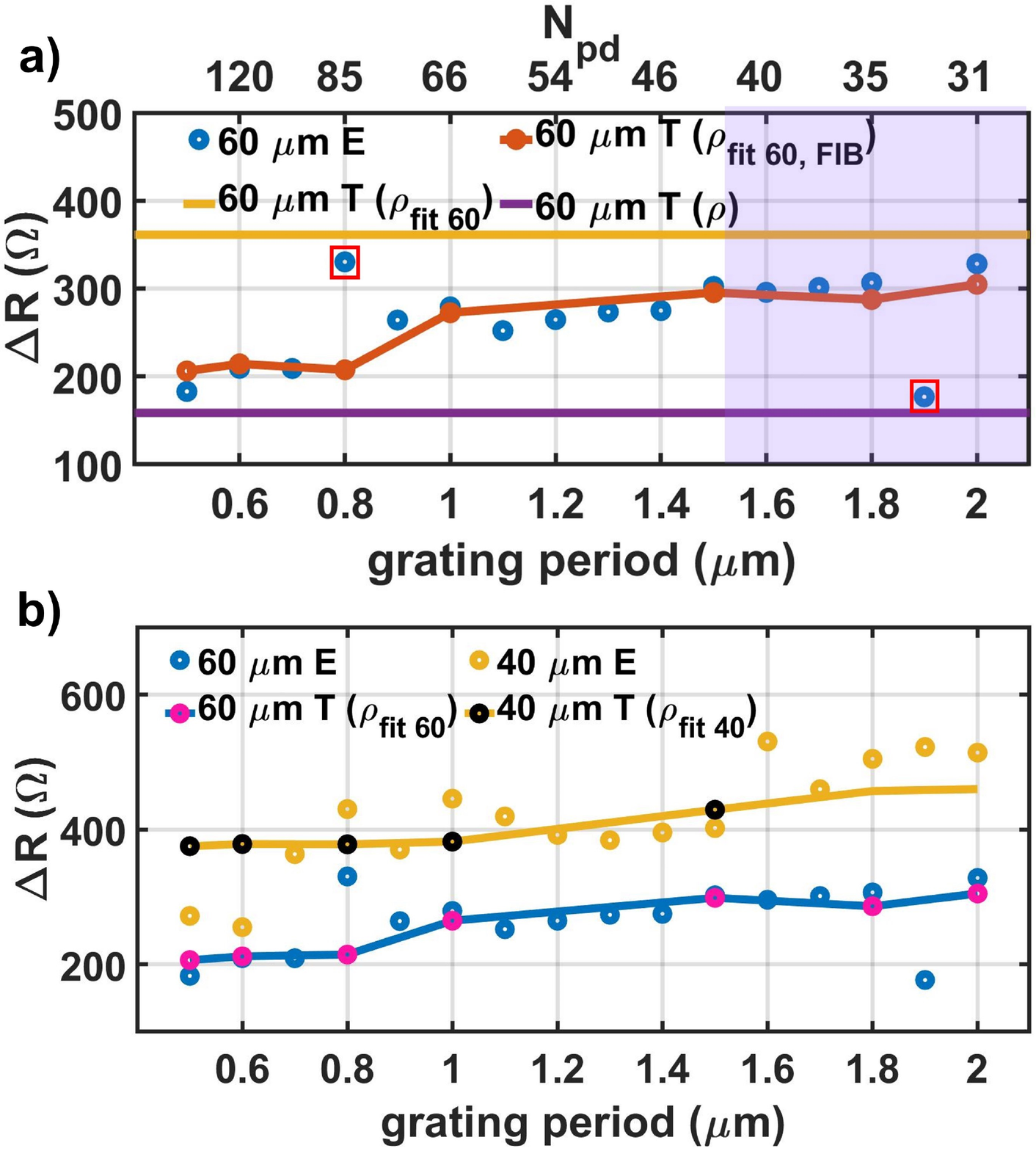}
\caption{ a) Resistance asymmetry ($\Delta$R, blue dots) plotted as a function of grating period and number of grating elements (\(\text{N}_{pd}\) = $N_{\Lambda}$/2) for $L_g$ = 60 \si{\um}.  We also show the theoretical estimates of ${\Delta}$R with (orange line) and without (yellow line) the FIB data from Fig.4(c). We use a $\rho_{fit 60}$ = 1.8x10$^{-3}$ $\Omega$-m as a free parameter in these fits. The theoretical fit based on $\rho$ = 7.9x10$^{-4}$ $\Omega$-m (purple line, no FIB data), corresponding to the foundry doping values is shown in purple. The shaded region of the plot indicates the period over which ellipsometric calibration of etch depths can be relied upon. We would like to note that the FIB-predicted ${\Delta}R$ curve (orange) includes the measured FF, $h$ and $\theta$ for each point, and is therefore non-monotonic unlike the $h$ data shown in Fig.4(c) (b) Measured $\Delta$R as a function of grating period for $L_g$ = 60 \si{\um} (blue dots) and 40 \si{\um} (yellow dots) devices. The theoretical fits, incorporating the FIB data are generated using, $\rho_{fit 60}$ = 1.8x10$^{-3}$ $\Omega$-m and $\rho_{fit 40}$ = 2.9x10$^{-3}$ $\Omega$-m, see text for further discussion. The FIB data points are explicitly shown in magenta and black dots for the $L_g$ = 60 \si{\um} and 40 \si{\um} devices respectively.} 
 \label{fig 5: expt theory FIB L60um}
\end{figure} 

To explore the quantitative agreement, we reconsider the data for the $L_g$ = 60 \si{\um} device shown in Fig.3. Fig.5(a) plots the resistance asymmetry (${\Delta}$R) with the blue dots showing the experimentally measured data, and the purple line showing the predicted (theory) ${\Delta}$R based on foundry doping estimated resistivity ($\rho$ = 7.9x10$^{-4}$ $\Omega$-m). For the theory, we consider $\rho$ to be a free parameter and use the large period data to achieve a best fit. Using this $\rho_{fit60}$ = 1.8x10$^{-3}$ $\Omega$-m and the measured values of the etch depth, fill factor and sidewall angle as a function of period from Fig.4(c), we can construct a better theoretical estimate of ${\Delta}$R which accounts for the etch micro-loading explicitly. The orange curve plots this result, with the FIB data points explicitly indicated. As can be seen, the improved predicted model (orange line) is a very good approximation to the measured ${\Delta}$R (blue dots). Taken in reverse the measured ${\Delta}$R can be used as a look-up table that maps to partial etch depths, and proves in principle that resistance asymmetries can be used to infer the geometric shapes in a non-destructive fashion. The yellow line shows the predicted ${\Delta}$R based on a modified resistivity ($\rho_{fit60}$= 1.8x10$^{-3}$ $\Omega$-m). The increase in perceived $\rho$ as compared to the value provided by foundry ($\rho$ = 7.9x10$^{-4}$ $\Omega$-m) could be due to a number of factors ranging from incomplete dopant activation to additional contact and interface resistance, which we discuss in detail below. 

While the data in Fig.5(a) shows good agreement with the model in principle, note that there is still some device to device resistance fluctuation, most clearly indicated by the presence of outlier data points at $\Lambda$ = 0.8 \si{\um} and 1.9 \si{\um}  (indicated by red boxes around these two points  in Fig.5(a)). We believe the main cause of this fluctuation is a change in contact (and interface) resistance between devices. One way to clearly see this effect is to look at the data in Fig.5(b) which plots ${\Delta}$R with respect to grating period for two sets of devices with lengths $L_g$ = 40 \si{\um} (yellow dots) and 60 \si{\um} (blue dots) respectively. The two datasets show roughly the same trend for ${\Delta}$R and the datasets can be reasonably fit using the FIB derived micro-loading curves (blue and yellow lines respectively). Here the FIB data ($h$, FF and $\theta$) is interpolated for $\Lambda$ $>$ 1.5 \si{\um} using a 3$^{rd}$ order polynomial. On the other hand, it is clear that the two curves are clearly offset with respect to each other and this is also confirmed by the best-fit $\rho$ values being different with $\rho_{fit 60}$ = 1.8x10$^{-3}$ $\Omega$-m and $\rho_{fit 40}$ = 2.9x10$^{-3}$ $\Omega$-m. The need to have two different $\rho$ values for different length devices makes it challenging from a metrological perspective, as one would naively expect the measured sample resistivity to be independent of $L_g$.

\begin{figure}[h!]
\centering\includegraphics[width=1\textwidth, trim = 0.5cm 3.5cm -0.5cm 3.8cm, clip]{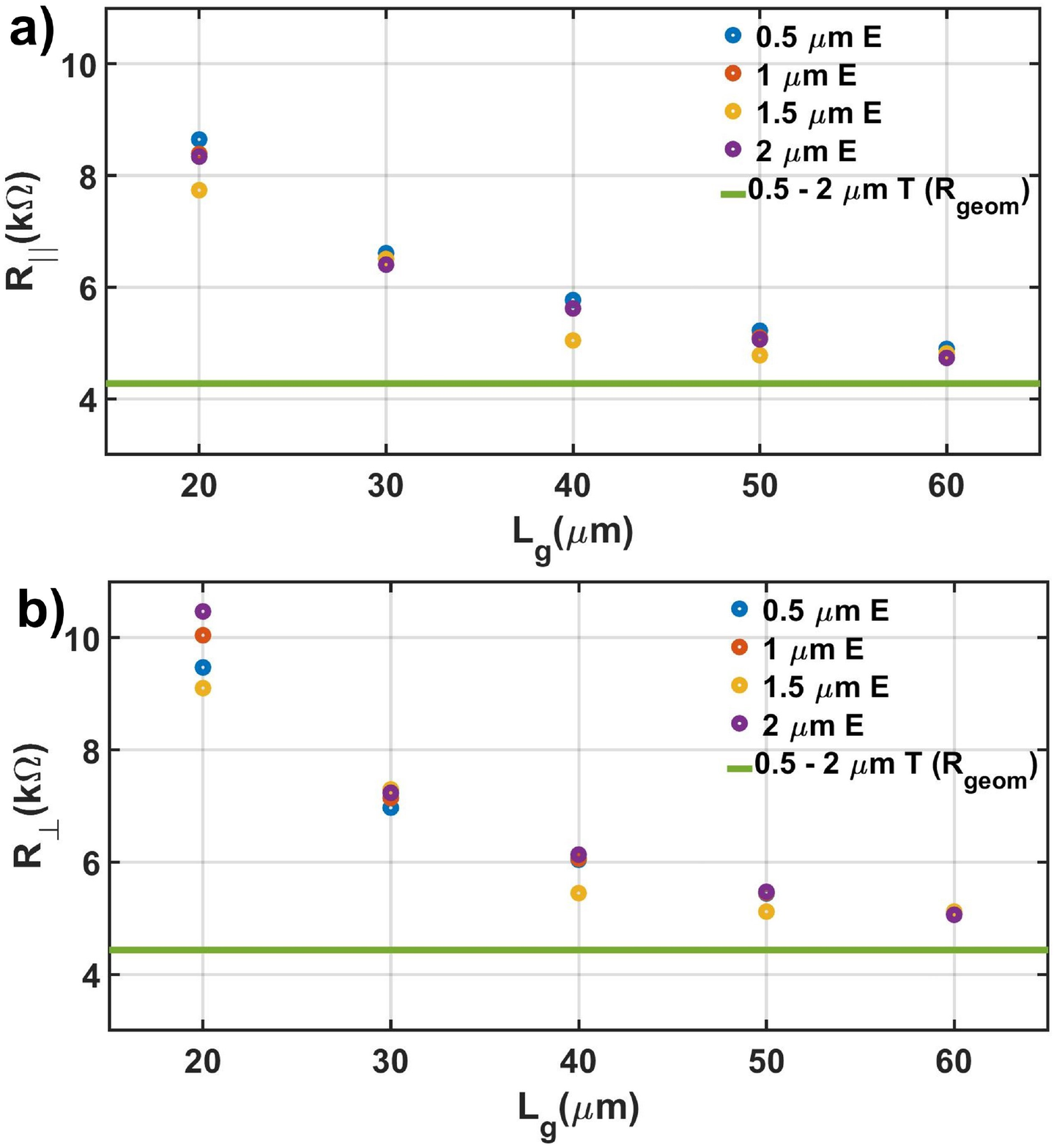}
\caption {Experimentally measured (E) and theoretically estimate (T) values of (a) \(\text{R}_{||}\) and b) \(\text{R}_{\perp}\) with respect to grating lengths for selected grating periods (0.5-2 \si{\um}). The theoretical values for both \(\text{R}_{||}\) and (b) \(\text{R}_{\perp}\) are estimated purely using the geometrical resistance (\(\text{R}_{geom}\)) and for a $\rho$ = 7.9x10$^{-4}$ $\Omega$-m, corresponding to the best-fit value obtained for the data in Fig.5(a)}
 \label{fig 6: R vs Lg}
\end{figure}

One way to understand the inferred $\rho(L_g)$ is to realize that our theoretical model only estimates the geometrical component of resistance (\(\text{R}_{geom}\)). More generally, the resistance of the device in either the parallel or perpendicular orientation can be written as:

\begin{equation}
    \mathrm R_{||,\perp} = 2* \mathrm R_{contact} + 2* \mathrm R_{interface} + \mathrm R_{geom}
\end{equation}

where $R_{contact}$ is the contact resistance at the metal semiconductor interface and $R_{interface}$ is the interface resistance between the contact region and cross-section of the gratings itself in both parallel and perpendicular directions. Inferring geometry from a resistance asymmetry is only valid if $R_{geom}\gg R_{contact},R_{interface}$. In addition, we have implicitly assumed that the $R_{contact}$ and $R_{interface}$ are nominally identical in the parallel and perpendicular orientations, and therefore cancel out in a ${\Delta}$R measurement. While this is mostly true for $R_{contact}$ due to symmetry, the $R_{interface}$ could be (and most likely is) slightly different in the two orientations giving rise to a systematic error. This is due to the fact that along the parallel direction of measurement, the interface between the contact pad and the grating is along the length of a grating element, while for the perpendicular direction, it is along the width of the grating elements and hence is more likely to get affected by the cross-section of the gratings or $R_{geom}$. To probe this, we can plot the individual $R_{||}$ and $R_{\perp}$ values for a given period as a function of grating length ($L_g$). Both the $R_{contact}$ and $R_{interface}$ values reduce as $L_g$ is increased while the $R_{geom}$ is unaffected by a change in $L_g$ (see eqns 6-7). One can then use this to estimate the nominal $L_g$ where the sample resistance is dominated by the geometrical resistance and therefore, ${\Delta}$R can be reliably used to infer the sample geometry and etch micro-loading.

The data is shown in Fig.6(a) for \(\text{R}_{||}\) and 6(b) for \(\text{R}_{\perp}\) for four different grating periods as a function of device length ($L_g$). One can see that both \(\text{R}_{||}\) and \(\text{R}_{\perp}\) decrease with increasing $L_g$ and converge towards the theoretical estimate based on the geometrical calculation, shown by the green line. The theoretical estimate uses the foundry provided doping estimated resistivity value ($\rho$ = 7.9x10$^{-4}$ $\Omega$-m) and one can see from these curves that the device length ($L_g$) needs to be  $>$ 60 \si{\um} to enable reliable reconstructions of geometry using ${\Delta}$R. This also agrees well with our main result from Fig.5(a), where the 60 \si{\um} data gave us the best reconstruction of micro-loading values with the sample resistivity closest to that provided by the foundry. We would like to note here that in these first-generation experiments, the maximum $L_g$ in our devices was 60 \si{\um} and this happens to lie right at the edge where the \(\text{R}_{geom}\) starts to dominate, and useful reconstruction can be achieved. In a sense, we found this result quite surprising as we expected this crossover to occur for smaller $L_g$ $\approx$ 30-40 \si{\um}.  Another trend that can be seen from both the \(\text{R}_{||}\) and \(\text{R}_{\perp}\) data is that the variance between the measured resistance values for different grating period devices of the same nominal sample length goes down as $L_g$ increases, leading to more reliable measurements. This can be explained by noting that there are more factors that can lead to fluctuations in contact and interface resistance, than the geometrical resistance and as one approaches the \(\text{R}_{geom}\) dominated regime, the spread in resistance values gets significantly tighter. Finally, note that the exact cross-over point for $R_{geom}$ dominated resistance will be set by the specific foundry process, in particular the doping levels available and the 60 \si{\um} value is specific to the foundry and doping process used in this work. On the other hand, measurements like those shown in Fig.6(a,b) can be used to reliably deduce the $L_{g}$ where the \(\text{R}_{geom}\) dominates over the contact and interface resistances.

\section{Outlook: Improvements to the method and extensions}

While in this work, we have clearly demonstrated that resistance asymmetries can be used to reconstruct the etch micro-loading curves of foundry processes, there are still some sources of error that need to be addressed to enhance the reconstruction accuracy. As discussed in the previous section, the accuracy improves as the \(\text{R}_{geom}\) dominates the measured resistance, and the 60 \si{\um} long devices were pretty much at the limit where acceptable reconstruction was feasible. Working with longer devices $L_g\approx$ 100 \si{\um} would reduce the reconstruction error due to the residual contact and interface resistances. Similarly, working with longer period ( > 2 \si{\um}) devices will help elucidate the spread in resistance across different grating periods and allow us to trace back the etch parameter reconstruction for shorter period devices.  A second issue that needs to be addressed is the observed device to device variation (see outliers indicated in Fig.5(a)). There are various potential sources for these, some of which are hard to predict, given the publicly available knowledge of the exact foundry process steps. These range from variations due to dopant implantation at an angle (7{\textdegree} in our case) and dopant implants through local oxide windows, which could lead to potential shadowing and micro-doping effects, i.e. variations in doping concentration with respect to feature size. Fig.5(a) represents the best dataset we could acquire from our measurements and while the general trends are observed even in shorter devices, as shown in Fig.5(b), the fluctuations are worsened, hinting at local micro-doping variations. 

Our method, relying on a subtraction, explicitly assumes that the contacts are as close to identical, post annealing, but empirically we have found this not to be the case, as indicated by the outlier data devices which were located right next to perfectly functioning contacts. In general, we find that outlier values of ${\Delta}$R correspond to bare \(\text{R}_{||}\) and \(\text{R}_{\perp}\) values that are nominally greater, and this is usually a clear signature that something other than the \(\text{R}_{geom}\) is dominating. In the SI, section S3, Fig.S2, we plot additional measured ${\Delta}$R data for shorter $L_g$ devices, and for the 20 \si{\um} devices in particular, the ${\Delta}$R fluctuations are primarily due to difference in contact and interface resistances and the geometrical fluctuations are completely masked out. As discussed above, even the 60 \si{\um} dataset lies at the edge of the region where the \(\text{R}_{geom}\) dominates. In the SI, section S4, Fig.S3 we have added data from another set of 60 \si{\um} devices from the same die as the data shown in Fig.5(a), and here the fluctuations in ${\Delta}$R make it challenging to infer geometry, especially if the fluctuations happen to occur in the small $\Lambda$ regime where the largest deviations are expected. On the other hand, as the data in Fig.5(a) shows, this is something that can be controlled and as long as a set of devices, all working in the right operating regime can be obtained, resistance asymmetries can be used to infer geometry.  
 
Given the complexity of a silicon photonic actives process, there will inevitably be some sources of error. The key question is how far can these errors be calibrated out. One of the main issues we had in these first generation devices was a lack of independent measurement of $\rho$. This could have been simply achieved by having an unetched silicon block (no gratings patterned) in the same geometry and verifying that the \(\text{R}_{||}\) and \(\text{R}_{\perp}\) were identical (the difference bounds our error). A lot of the other systematic effects (contact resistance variation, micro-doping, dopant non-uniformity with thickness) can be eliminated through statistics. The micro-doping and shadowing effects can be removed, for instance, by having gratings at various in-plane orientations. Note that while averaging ${\Delta}$R across multiple identical device datasets can in principle reduce fluctuations, it is valid only if the bare \(\text{R}_{||}\) and \(\text{R}_{\perp}\) values are comparable, as discussed further in the SI, section S4. 

Moving forward, this method can be extended in several directions. Generalizing a resistance asymmetry to an impedance asymmetry would allow this method to be extended to non-conductive photonics platforms like silicon nitride, although this would involve significant added complexity due to the need to account for fringing fields \cite{pozar2000microwave}. Even within silicon, while this work has focused on linear rectangular gratings, the ${\Delta}$R can be generalized to reconstruct more complex elliptical shapes such as encountered in photonic crystals or waveguide micro-ring resonator coupling. Figuring out a way to extend these two port measurments to 4-port geometries to eliminate the \(\text{R}_{contact}\) and \(\text{R}_{interface}\) would also help to improve the accuracy of this method. Finally, one of our key goals for developing this approach is that, by monitoring ${\Delta}$R inside an etch chamber, one can in principle reconstruct the partial etch depth with real-time accuracy and stop the etch with high precision at the desired etch depth, something that is currently challenging to do with other metrology methods.

\section{Conclusions}
In this work, we have used electrical resistance asymmetries (${\Delta}$R) to extract the geometry of Si nano gratings by using a two probe electrical resistance measurement. ${\Delta}$R monotonically increases with respect to grating periods and tends to saturate at longer periods ($\Lambda$ > 1.5 \si{\um})  which is attributed to the micro-loading effect of the dry etching process. We have used this method to demonstrate the reconstruction of a foundry etch micro-loading curve. This was confirmed with FIB cross-section of the devices. It was found that the side-wall angle had relatively little effect on the ${\Delta}$R, whereas it is extremely sensitive to both etch depth and fill factor. Both $R_{||}$ and $R_{\perp}$ decreases with grating lengths and starts to saturates for lengths where $R_{geom} \gg R_{contact},R_{interface}$, which is found to be 60 \si{\um} set by the foundry process used in this research. Operating in this geometrical resistance dominated regime is critical for this method to work. Some sources of error, mainly resistance fluctation between nominally identical devices, need to be addressed to improve the overall reconstruction accuracy and make this method suitable for nanofabrication metrology.

\section{Funding}
We would like to thank the European Research Council (ERC-StG SBS 3-5, 758843) and the UK Engineering and Physical Sciences Research Council (EP/V052179/1) for funding support.  Nanofabrication was carried out using equipment funded by an EPSRC captial equipment grant
(EP/N015126/1). The foundry chip fabrication was carried out as part of the Cornerstone project, funded by UK EPSRC (EP/L021129/1).

\section{Acknowledgments}
 The authors would like to thank Callum Littlejohns from Cornerstone for help with the chip tapeout. We would also like to thank Henkjan Gersen, Ian Lindsay, Karen Grutter, Kartik Srinivasan, and Rob Ilic for valuable discussions and suggestions. 

\beginsupplement

\section{Derivation of (\(\text{R}_{||}\) and \(\text{R}_{\perp}\)) components of SOI grating }
Here, we provide the derivation of the results stated in eqns. (6) and (7) in the main text. Our notation below follows the labelling convention shown in Figs.2(c) and 2(d).

We model the resistance of the grating structures by breaking them into component sections, denoted in Fig.2(c), and summing up the overall contribution. In general, the electrical resistance of any 'wire-like' geometry, assuming uniform current flow, is given by:
\begin{gather}
\mathrm R = \frac{\rho L} {A}
\end{gather}
where $\rho$ is the material (silicon) resistivity, $L$ is the length and $A$ is the cross-sectional area of the element. To calculate the \(\text{R}_{||}\) component in Fig.2(c), the length (long-axis) of the grating is normal to the page and $A$ is calculated individually for cross-sectional area sections 1, 2 and 3 respectively.
\begin{gather}
\mathrm R_{||1}= \frac{\rho L_g} {(H-h) (W - \mathrm {tan}\theta)} \\
\mathrm R_{||2}= \frac{\rho L_g} {(H-h) (h\mathrm {tan}\theta) + (\frac{1}{2} \mathrm{tan}\theta)}\\
\mathrm R_{||3} = \frac{\rho L_g}{Hw}
\end{gather}

Given the current flows in parallel in the three sections, the total contribution towards \(\text{R}_{||}\) becomes:
\begin{gather}
\frac{1}{\mathrm R_{||}} = \frac{1}{\mathrm R_{||1}} + \frac{1}{\mathrm R_{||2}} + \frac{1}{\mathrm R_{||3}} \\
\frac{1}{\mathrm R_{||}} = \frac {1}{\rho L_g} (HW - hW + hw + \frac{1}{2} h^2 \mathrm{tan}\theta) \\
\mathrm R_{||}\ = \frac{\rho L_g}{(HW - hW + hw + \frac{1}{2}h^2 \mathrm{tan}\theta)}
\end{gather}

Similarly, for estimating the perpendicular component (\(\text{R}_{\perp}\)), $L$ of the grating is now along the cross-section ($W$) of the device, as indicated in Fig.2(c). Similar to the \(\text{R}_{||}\) case, we can evaluate the resistance contributions of each section. Region 2 in this case requires an integral given the change in cross-section, as indicated in Fig.2(d). 
\begin{gather}
\mathrm R_{\perp1}\ = \frac{\rho (W - h \mathrm{tan}\theta - w)}{L_g (H-h)}    
\end{gather}
For region 2, we integrate over the elemental resistance contributions as indicated in Fig.2(d):
\begin{gather} 
 d \mathrm R_{\perp2}\ = \int_0^{h\mathrm{tan}\theta} \! \frac{\rho \mathrm{d}x}{L_g (H-h + \frac{x}{\mathrm{tan}\theta})} \\
 \mathrm R_{\perp2}\ = \frac{\rho}{L_g} \mathrm {tan}\theta  ln\left(\frac{H}{H-h}\right)\\
 \mathrm R_{\perp3}\ = \frac{\rho w}{L_g H} 
 \end{gather}
 The total contribution for \(\text{R}_{\perp}\) can be calculated by noting that the elemental contributions now occur in series:
 \begin{gather}
 \mathrm R_\perp\ =  \mathrm R_{\perp1}\ + \mathrm R_{\perp2}\ + \mathrm R_{\perp3}\ \\
 \mathrm R_\perp\ = \frac{\rho}{L_g} \left( \frac{W-h \mathrm {tan}\theta-w}{H-h} + \mathrm {tan}\theta  \ln\left(\frac{H}{H-h}\right) +\frac{w}{H} \right) 
\end{gather}

\section{Sensitivity of ${\Delta}$R to changes in $h$ and $\theta$.}
Here we provide the equations used to plot the first derivative of \(\text{R}_{||}\) and \(\text{R}_{\perp}\) with respect to $h$ and $\theta$, to plot sensitivity of $\Delta$R with respect to $h$ and $\theta$:

\begin{gather}
\frac{\mathrm d\mathrm R_{||}}{\mathrm dh}= -\rho L_g \frac{h \mathrm{tan}\theta  +w-W} {(HW-hW+hw+ \mathrm{tan}\theta \frac{h^2}{2})} \\
\frac{\mathrm d\mathrm R_{||}}{\mathrm d \theta}= -\frac{\rho L_g h^2 \mathrm{sec^2}\theta}{2(HW-hW+hw+ \mathrm{tan}\theta \frac{h^2}{2})} \\
\frac{\mathrm d\mathrm R_{\perp}}{\mathrm dh}= \frac{\rho}{L_g}  \frac{(-h \mathrm{tan}\theta -w+W)} {(H-h)^2} \\
\frac{\mathrm d\mathrm R_{\perp}}{\mathrm d \theta}= -\frac{\rho}{L_g} ln(\frac{H}{H-h}) \mathrm {sec^2}\theta -\frac{h \mathrm {sec^2}\theta}{H-h}
\end{gather}
Gradient of $\Delta$R is then given by the following equations and is plotted below in Fig.S1:
\begin{gather}
\mathrm d \frac {\Delta \mathrm R} {\mathrm dh}= \frac{\mathrm d\mathrm R_{\perp}}{\mathrm dh} - \frac{\mathrm d\mathrm R_{||}}{\mathrm dh}\\
\mathrm d \frac {\Delta \mathrm R} {\mathrm d\theta}= \frac{\mathrm d\mathrm R_{\perp}}{\mathrm d\theta} - \frac{\mathrm d\mathrm R_{||}}{\mathrm d\theta}
\end{gather}

\begin{figure}[H]
\centering\includegraphics[width=1\textwidth, trim = 0.8cm 3.3cm 0.2cm 3.5cm, clip]{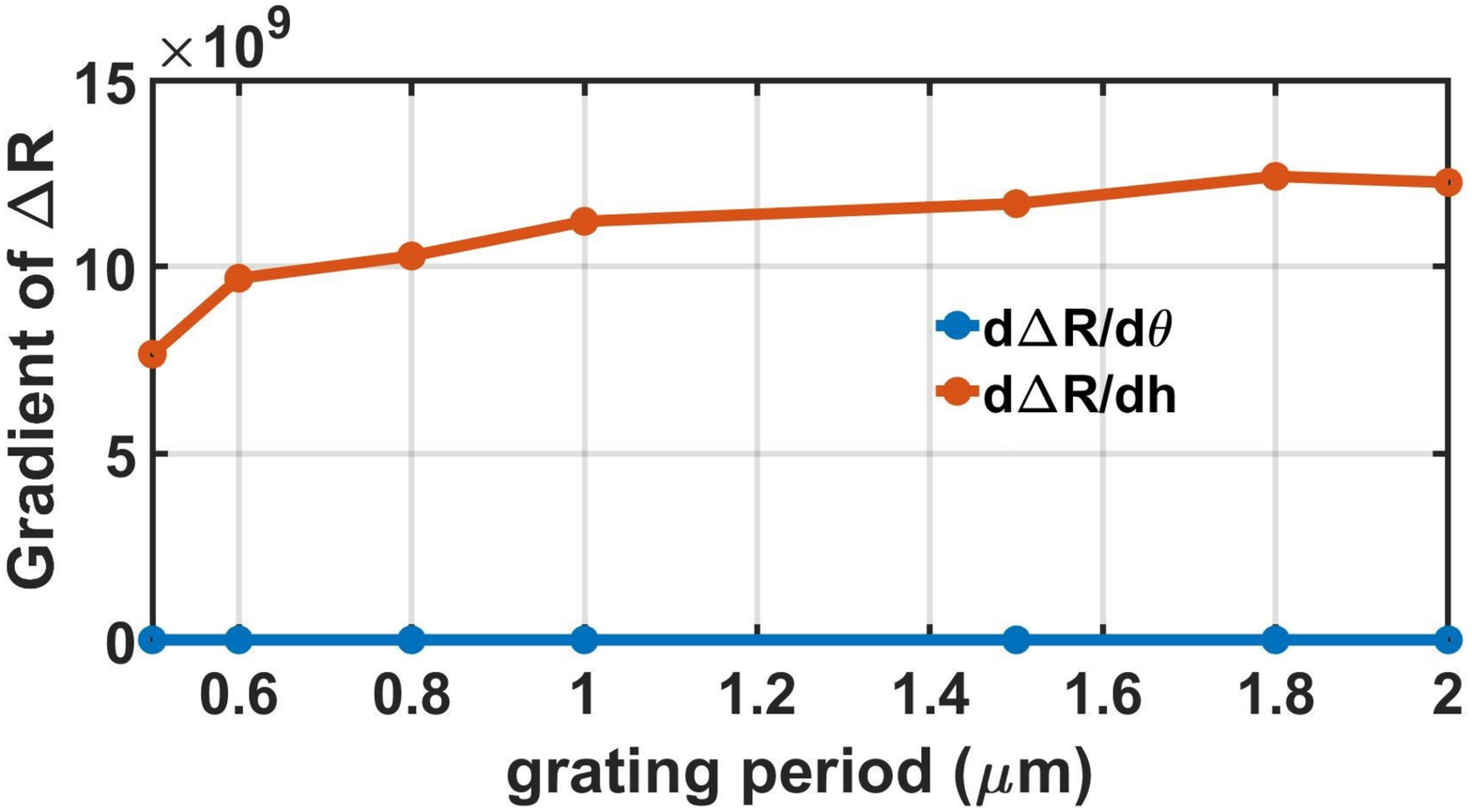}
\caption {Resistance asymmetry ($\Delta$R) as a function of grating period for grating length of 60 \si{\um} measured experimentally (E) for derivative of $\Delta$R with respect to etch depth ($h$) and side wall angle ($\theta$) measured by FIB.}
 \label{fig S1: derivative of delta R wrt h and theta for L60}
\end{figure}

\section{Additional $L_g$ datasets for measured resistance asymmetry ($\Delta$R)}
\begin{figure}[H]
\centering\includegraphics[width=1\textwidth, trim = 0.5cm 3.25cm 0.5cm 4.0cm, clip]{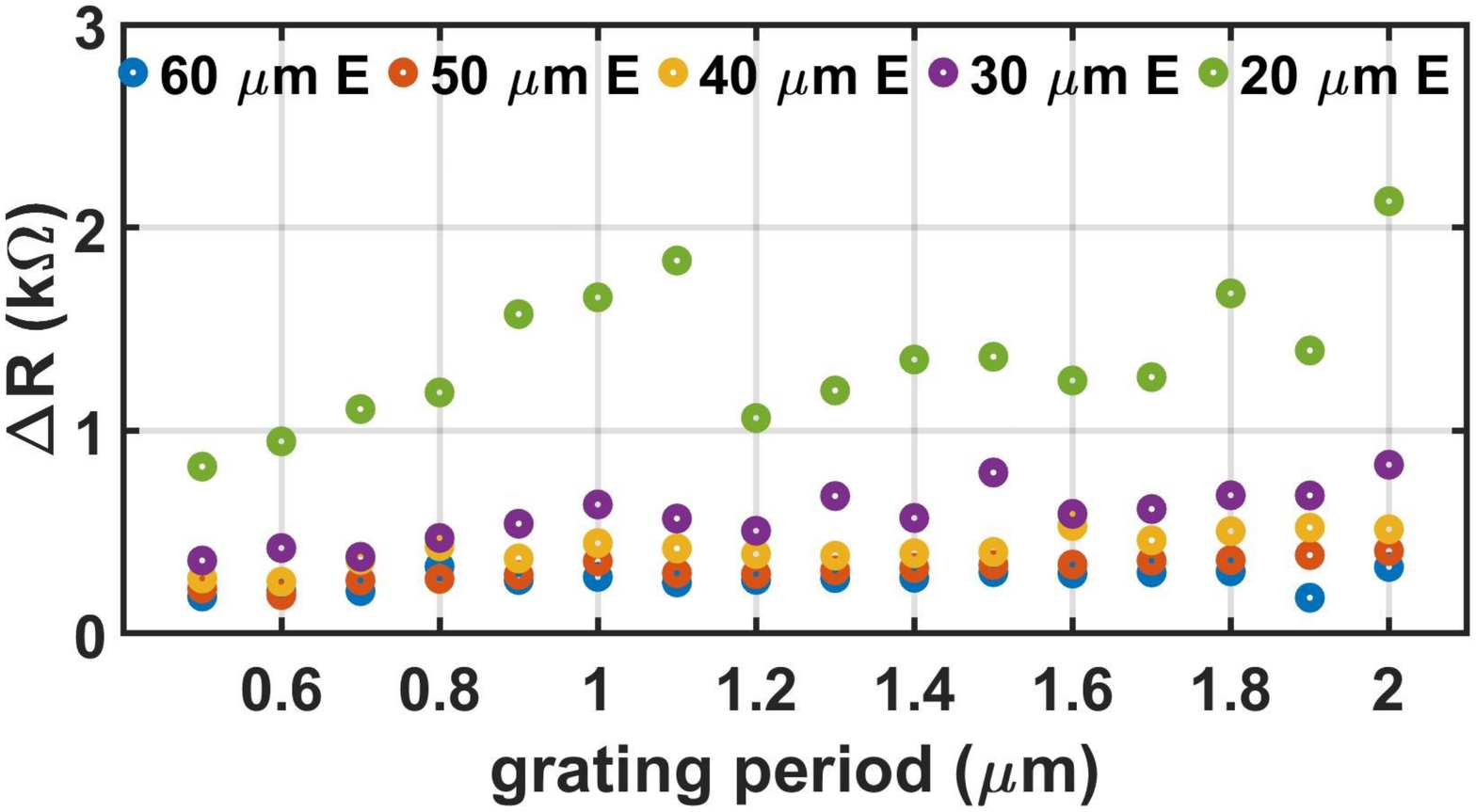}
\caption {Resistance asymmetry ($\Delta$R) as a function of grating period for grating length of 60 \si{\um}  to 20 \si{\um} measured experimentally (E). As noted in the main text, for short $L_g$, the device resistance has significant contributions from interface and contact resistances, and fluctuations in these end up reflecting in the differential resistance measurement. This is particularly clear in the data for the 20 \si{\um} devices, where the fluctuations in ${\Delta}R$ due to contact and interface resistance values mask any changes due to the geometrical resistance.}
 \label{fig S2: expt delta R for L60 to L20}
\end{figure}

\section{Resistance asymmetry ($\Delta$R) as a function of grating period for grating length of 60 \si{\um} for 2 sets of devices. }

\begin{figure}[H]
\centering\includegraphics[width=1\textwidth, trim = 0.2cm 3.3cm -1.4cm 3.5cm, clip]{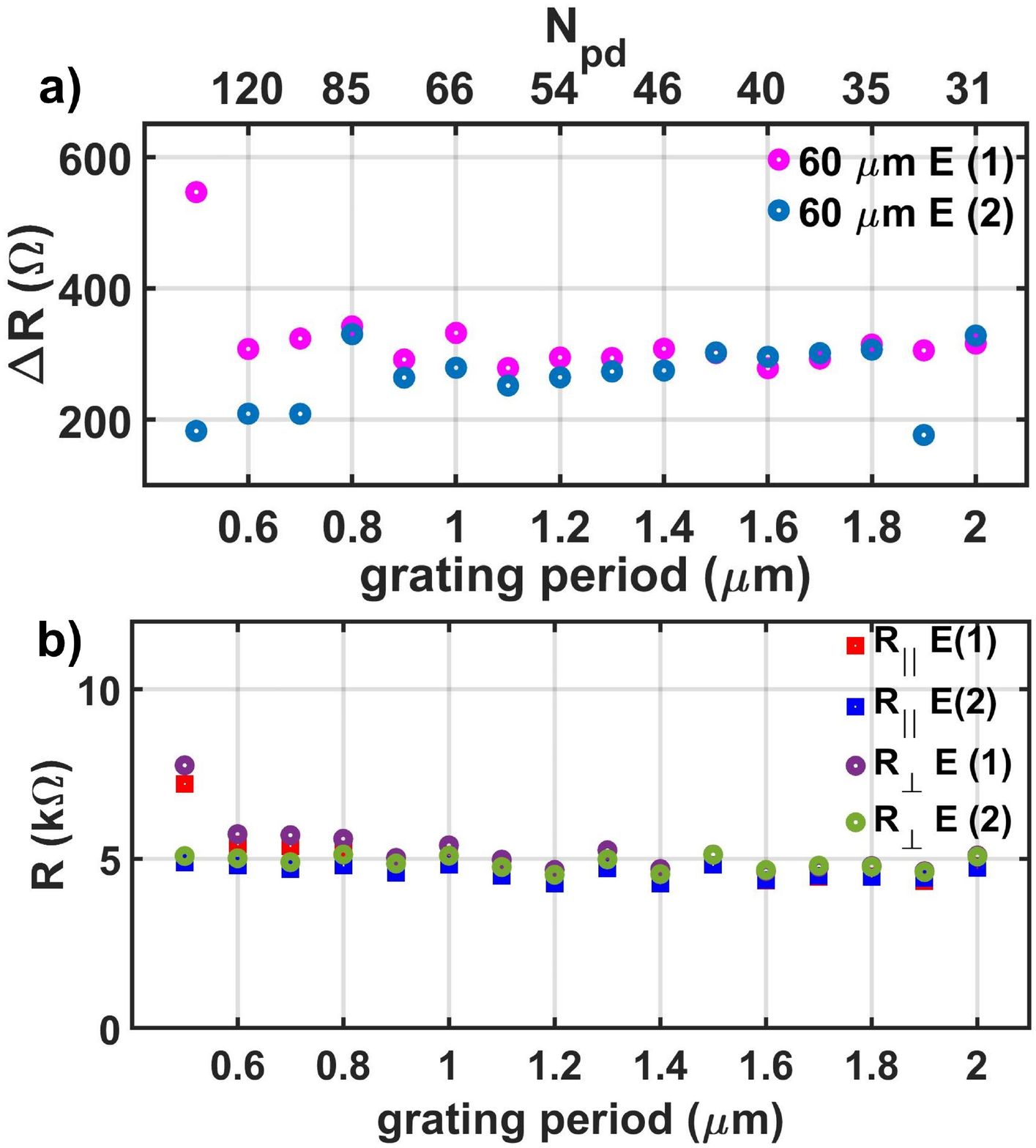}
\caption {(a) Resistance asymmetry ($\Delta$R) as a function of grating period for grating length of 60 \si{\um} measured experimentally (E) for two sets of devices. (b) \(\text{R}_{||}\) and \(\text{R}_{\perp}\) with respect to grating period measured experimentally for grating length 60 \si{\um} for 2 datasets. E (2)) represents the dataset used in the main manuscript (Fig. 3 and Fig. 5) for length of 60 \si{\um} and E (1)) are the dataset measured on another set of devices for the grating length of 60 \si{\um}.}
\label{fig S3: expt delta R for 2 sets for L60}
\end{figure}


\bibliography{References}

\end{document}